# Non-Linear Speech coding with MLP, RBF and Elman based prediction[1]

Marcos Faúndez-Zanuy

Escola Universitària Politècnica de Mataró
Universitat Politècnica de Catalunya (UPC)
Avda. Puig i Cadafalch 101-111, E-08303 Mataró (BARCELONA) SPAIN
`faundez@eupmt.es`

**Abstract.** In this paper we propose a nonlinear scalar predictor based on a combination of Multi Layer Perceptron, Radial Basis Functions and Elman networks. This system is applied to speech coding in an ADPCM backward scheme. The combination of this predictors improves the results of one predictor alone. A comparative study of this three neural networks for speech prediction is also presented.

## 1. Introduction

Time series analysis and prediction has potential applications in several fields, such as automation and quality control, financial time series analysis, stock exchange, efficient planning and production, operator assistance in process industry, medicine, weather, etc. One important application of time series prediction is found in speech signals related applications. For instance, most of the speech coders use some kind of prediction. The most popular one is the scalar linear prediction, but several papers have shown that a nonlinear predictor can outperform the classical LPC linear prediction scheme [1-3].

In our previous work, we used a Multi Layer Perceptron (MLP) instead of the classical linear predictor, for speech coding purposes. In order to keep the speech coder stable, it was introduced in a closed loop scheme with a quantizer, named ADPCM (Adaptive differential PCM).

In this paper, we study two new different neural networks predictors (Elman recurrent network and Radial Basis Functions), that replace and combine with our scheme proposed in [1]. Figure 1 shows the scheme of the ADPCM speech encoder. The neural predictor is updated on a frame basis, using a backward strategy. That is, the coefficients are computed over the previous frame. Thus, it is not needed to transmit the coefficients of the predictor, because the receiver has already decoded the previous frame and can obtain the same set of coefficients.

---

[1] This work has been supported by the CICYT TIC2000-1669-C04-02 and COST-277

This paper shows that the combination of this three kind of neural net predictors can improve the results of one predictor alone and can reduce the computational burden of the original ADPCM scheme with MLP prediction that we have used in our previous work.

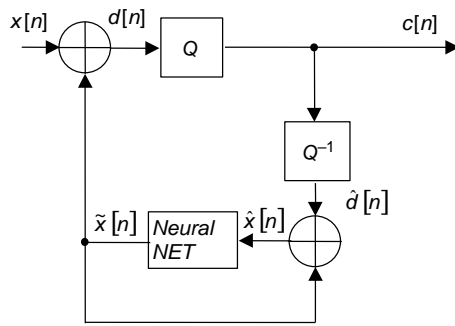

**Fig. 1.** ADPCM scheme with neural net prediction

## 2. Conditions of the experiments

This section describes the conditions of the experiments.

### 2.1 Conditions of the experiments

The experimental results have been obtained with an ADPCM speech coder with an adaptive scalar quantizer based on multipliers [4]. The number of quantization bits is variable between Nq=2 and Nq=5, that correspond to 16kbps and 40kbps (the sampling rate of the speech signal is 8kHz). We have encoded eight sentences uttered by eight different speakers (4 males and 4 females). These are the same sentences that we used in our previous work [1-3].

### 2.2 Evaluation of the results

For waveform speech coders, we can evaluate the speech encoder quality using the Segmental signal to Noise Ratio (SEGSNR). The SEGSNR is computed with the expression $SEGSNR = \frac{1}{K}\sum_{j=1}^{K} SNR_j$ , where $SNR_j$ is the signal to noise ratio (dB) of frame $j$ : $SNR = \frac{E\{x^2[n]\}}{E\{e^2[n]\}}$, and $K$ is the number of frames of the encoded file.

## 3. MLP, Elman, and RBF networks parameter settings.

In this section we describe the new prediction networks and their parameter setting, with special emphasis on Elman and RBF networks.

### 3.1 Multi Layer Perceptron

We have used the same adjustments for the MLP than in our previous work:
- We fixed the structure of the neural net to 10 inputs, 2 neurons in the hidden layer, and one output.
- The selected training algorithm was the Levenberg-Marquardt, that computes the approximate Hessian matrix, because it is faster and achieves better results than the classical backpropagation algorithm.
- We also applied a multi-start algorithm with five random initializations for each neural net, and a committee between these five networks [3].

The combination between Bayesian regularization with a committee of neural nets increased the SEGSNR up to 1.2 dB over the MLP trained with the Levenberg-Marquardt algorithm [5-6], and decreases the variance of the SEGSNR between frames. For more information about the MLP setup you can refer to [1-3]. Anyway, this study has been made with the neural network toolbox of MATLAB 6.5, that uses a different random initialization algorithm than previous versions, so there are small differences of SEGNSR than the previous reported results for MLP.

### 3.2 Elman network

The Elman network commonly is a two-layer network with feedback from the first-layer output to the first layer input. The Elman network has *tansig* neurons in its hidden (recurrent) layer, and *linear* transfer functions in its output layer. This combination is special in that two-layer networks with these transfer functions can approximate any function (with a finite number of discontinuities) with arbitrary accuracy. The only requirement is that the hidden layer must have enough neurons. More hidden neurons are needed as the function being fitted increases in complexity. Note that the Elman network differs from conventional two-layer networks in that the first layer has a recurrent connection. The delay in this connection stores values from the previous time step, which can be used in the current time step. Thus, even if two Elman networks, with the same weights and biases, are given identical inputs at a given time step, their outputs can be different due to different feedback states.

In this paper we have used the Elman network with Bayesian Regularization and the Levenberg-Marquardt algorithm in a similar fashion and parameter setting than the MLP. Figure 2 shows a comparison between MLP and Elman networks architecture.

One important parameter setting is the number of epochs. We have evaluated two cases: 6 and 50 epochs. These are the same values that we used in our previous work for the MLP.

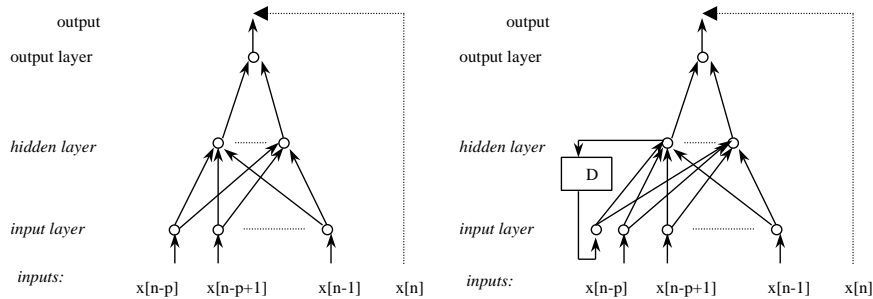

**Fig. 2.** MLP and Elman networks

### 3.3 Radial Basis Function

While Elman networks are close together to MLP, the RBF networks may require more neurons than MLP or Elman networks, but they can be fitted to the training data with less time. On the other hand, the transfer function is different:

$$radbas[n] = e^{-n^2}$$

The RBF network consists on a Radial Basis layer of $S$ neurons and an output linear layer. The output of $i^{th}$ Radial Basis neuron is $R_i = radbas(\|\vec{w}_i - \vec{x}\| \times b_i)$, where:

- $\vec{x}$ is the $p$ dimensional input vector
- $b_i$ is the scalar bias or spread ($\sigma$) of the gaussian
- $\vec{w}_i$ is the $p$ dimensional weight vector of the Radial Basis neuron $i$.

In our case, the output is just one neuron. Figure 3 shows the scheme of a RBF network.

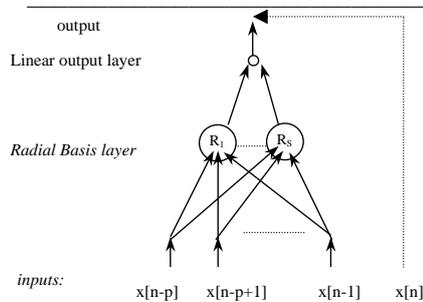

**Fig. 3.** RBF network architecture

The radial basis function has a maximum of 1 when its input is 0. As the distance between w and p decreases, the output increases. Thus, a radial basis neuron acts as a detector that produces 1 whenever the input $\vec{x}$ is identical to its weight vector $\vec{w}_i$. The bias *b* allows the sensitivity of the *radbas* neuron to be adjusted. For example, if a neuron had a bias of 0.1 it would output 0.5 for any input vector $\vec{x}$ at vector distance of 8.326 (0.8326/b) from its weight vector $\vec{w}_i$, because $e^{-0.8326^2} = 0.5$.

We have studied the relevance of two parameters: spread and number of neurons. First, we have evaluated the SEGSNR as function of the spread of the gaussian functions. Figure 4, on the left, shows the results using one sentence, for spread values ranging 0.011 to 0.5 with an step of 0.01 and *S*=50 neurons. It also shows a polynomial interpolation of third order, with the aim to smooth the results. Based on this plot, we have chosen a spread value of 0.22. Using this value, we have evaluated the relevance of the number of neurons. Figure 4, on the right, shows the results using one sentence and a number of neurons ranging from 5 to 100 with an step of 5. This plot also shows an interpolation using a third order polynomial. Using this plot we have chosen an RBF architecture with *S*=20 neurons. If the number of neurons (and/ or the spread of the guassians) is increased, there is an overfit (over parameterization that implies a memorization of the data and a loose of the generalization capability).

Radial basis neurons with weight vectors quite different from the input vector $\vec{x}$ have outputs near zero. These small outputs have only a negligible effect on the linear output neurons. In contrast, a radial basis neuron with a weight vector close to the input vector $\vec{x}$ produces a value near 1. If a neuron has an output of 1 its output weights in the second layer pass their values to the linear neurons in the second layer. In fact, if only one radial basis neuron had an output of 1, and all others had outputs of 0's (or very close to 0), the output of the linear layer would be the active neuron's output weights. This would, however, be an extreme case. Each neuron's weighted input is the distance between the input vector and its weight vector. Each neuron's net input is the element-by-element product of its weighted input with its bias.

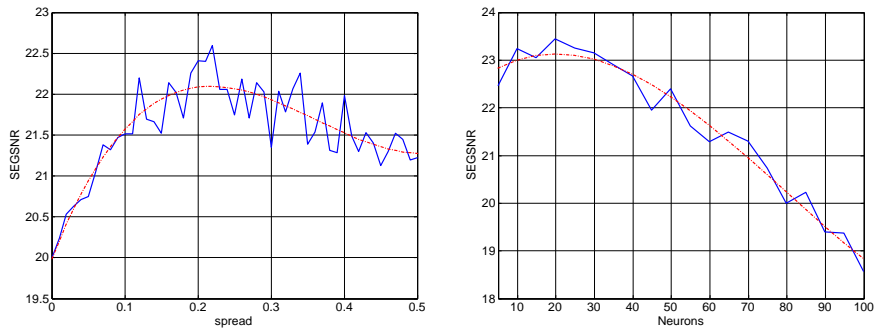

**Fig. 4.** Relevance of the σ of the gaussians;   Relevance of the number of neurons, with σ=0.22

The algorithm for training the RBF is the following:
- The algorithm iteratively creates a radial basis network one neuron at a time. Neurons are added to the network until the maximum number of neurons has been reached.
- At each iteration the input vector that results in lowering the network error the most, is used to create a radial basis neuron.

This problem of over/under fit can also be understood trying to interpolate between samples of a one dimensional signal using a RBF. Figure 5 shows several examples of gaussians, signal to fit, and output of the RBF for training samples and interpolated samples. It is interesting to observe that the output of the RBF is zero is those parts not covered by any gaussian (around ± 0.5 in the first example with 10 gaussians).

σ=0.06 and 10 gaussians

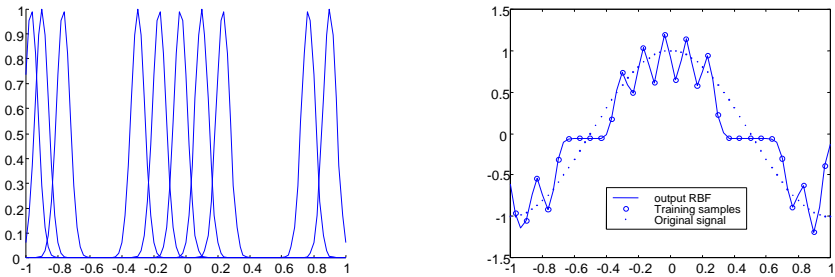

σ=0.3 and 3 gaussians

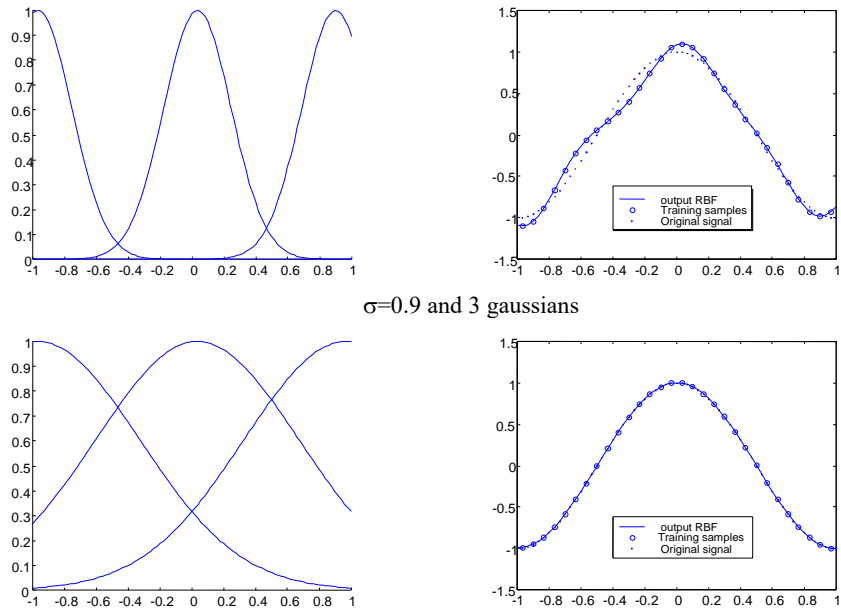

σ=0.9 and 3 gaussians

**Fig. 5.** Example of function approximation using RBF with different settings.

## 4. Results

This section describes the results using one neural net predictor and the combination between the three different kinds of neural net predictors.

Table 1 shows the results using one single kind of neural net predictor and different parameters. For instance, third column corresponds to a committee of five MLP (one different random initialization per network), and each net trained with 6 epochs. Table 2 shows the results for the combined system.

**Table 1.** Mean (m) and standard deviation (σ) of the SEGSNR for several predictors and quantization bits (Nq)

|    | 1 MLP 6 epoch | | 1 MLP 50 epoch | | 5 MLP 6 epoch | | 5 MLP 50 epoch | | 5 ELMAN 6 epoch | | 5 ELMAN 50 epoch | | 1 RBF | |
|----|------|-----|-------|-----|-------|-----|-------|-----|-------|-----|-------|-----|-------|-----|
| Nq | m | σ | m | σ | m | σ | m | σ | m | σ | m | σ | m | σ |
| 2  | 11.29 | 5.8 | 13.11 | 7.6 | 12.42 | 6.5 | 14.34 | 6.6 | 12.56 | 6.4 | 13.60 | 7.0 | 11.65 | 7.7 |
| 3  | 16.83 | 7.1 | 20.13 | 7.5 | 18.74 | 5.9 | 20.70 | 7.7 | 18.59 | 6.3 | 20.14 | 7.9 | 18.40 | 6.6 |
| 4  | 22.22 | 6.0 | 25.52 | 7.9 | 23.79 | 5.9 | 26.07 | 8.2 | 23.73 | 6.2 | 25.25 | 7.9 | 23.69 | 6.1 |
| 5  | 27.12 | 6.0 | 30.23 | 8.1 | 28.39 | 6.5 | 30.9  | 7.9 | 28.59 | 6.2 | 30.27 | 8.3 | 28.22 | 6.3 |

**Table 2.** Mean and standard deviation of the SEGSNR for several combinations

| Nq | RBF+MLP+ELM Mean 6 epoch | | RBF+MLP+ELM Median 6 epoch | | RBF+MLP+ELM Mean, 50 epoch | | RBF+MLP+ELM Median, 50 epoch | |
|---|---|---|---|---|---|---|---|---|
| | m | σ | m | σ | m | σ | m | σ |
| 2 | 12.65 | 6.3 | 12.91 | 5.4 | 13.74 | 6.9 | 14.05 | 6.4 |
| 3 | 19.05 | 5.8 | 18.71 | 6.4 | 20.25 | 7.3 | 20.62 | 7.3 |
| 4 | 24.04 | 6.2 | 23.76 | 6.1 | 25.33 | 7.3 | 25.97 | 7.1 |
| 5 | 28.85 | 6.0 | 28.41 | 6.3 | 30.01 | 8.0 | 30.87 | 7.4 |

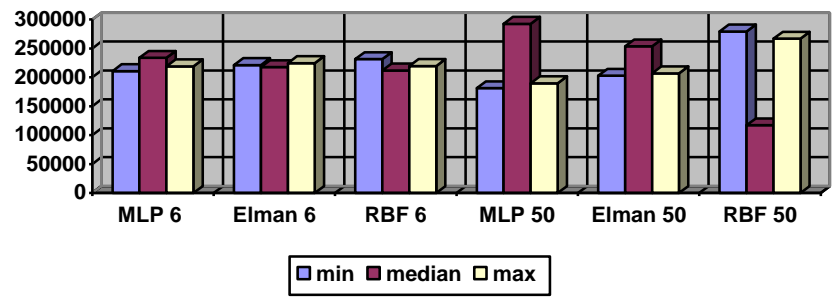

**Fig. 6.** Number of frames with minimum, median and maximum output for each predictor

For the combined scheme with MLP+ Elman+ RBF, all the predictors run in parallel for each sample, and two different combination strategies have been used: mean and median of the three outputs. Figure 6 shows the number of frames with minimum, median and maximum predicted value for each predictor, after sorting the three outputs for each sample. These results have been obtained with 6 and 50 epochs and median combination between the three outputs. For the RBF, the number of epochs has no sense. Thus RBF 6 means that the RBF network has been used in conjunction with MLP and Elman trained with 6 epochs. It is interesting to observe that the "best" predictor (from table 1 it can be deduced that the best predictor alone is the MLP) tends to be always in the middle between RBF (that tends to give smaller values) and Elman (that tends to give higher values).

## 5. Conclusions

In this paper we have evaluated three different kinds of neural networks for speech coding: Multi Layer Perceptron, Elman, and RBF. The comparison between them has shown the following:
- There are few differences in SEGSNR when using just one kind of predictor, although the MLP and Elman network can outperform the RBF when the number of epochs is 50.

Faundez-Zanuy, M. (2003). Non-Linear Speech coding with MLP, RBF and Elman based prediction[1] . In: Mira, J., Álvarez, J.R. (eds) Artificial Neural Nets Problem Solving Methods. IWANN 2003. Lecture Notes in Computer Science, vol 2687. Springer, Berlin, Heidelberg. https://doi.org/10.1007/3-540-44869-1_85
- The combination of the three kind of neural predictors yields an improvement in SEGSNR. This is equivalent to a committee of experts in the field of pattern recognition (classification), where the combination of different classifiers can outperform the results obtained with one single classifier.

The combination of several predictors is similar to the Committee machines strategy [7]. If the combination of experts were replaced by a single neural network with a large number of adjustable parameters, the training time for such a large network is likely to be longer than for the case of a set of experts trained in parallel. The expectation is that the differently trained experts converge to different local minima on the error surface, and overall performance is improved by combining the outputs of each predictor.